\documentclass[onecollarge,natbib]{svjour2}
\bibpunct{[}{]}{;}{n}{}{,} 
\def\scr{\rm\scriptscriptstyle }
\smartqed  
\usepackage{graphicx}
\journalname{Archive of Applied Mechanics}
\begin{document}

\title{Disappearance of Mott oscillations in sub-barrier elastic scattering of identical nuclei and atomic ions \thanks{The authors would like to thank CNPq, CAPES, FAPESP and FAPERJ for the partial financial support.}
}


\author{M. S. Hussein \and L. F. Canto \and R. Donangelo \and W. Mittig}

\authorrunning{Short form of author list} 

\institute{M. S. Hussein \at
               Instituto de F\'{i}sica and Instituto de Estudos Avan\c{c}ados, Universidade de S\~{a}o Paulo, S\~{a}o Paulo, 
               Brazil\\  Departamento de F\'{i}sica, Instituto Tecnol\'{o}gico de Aeron\'{a}utica, CTA, S\~{a}o Jos\'{e} dos Campos, 
               S.P., Brazil \\ 
               \email{hussein@if.usp.br}
                \and
                L. F. Canto \at
                Instituto de F\'isica, Universidade Federal do Rio de Janeiro, C.P. 68528, Rio de Janeiro, R.J., 21941-972, Brazil \\
                Instituto de F\'isica, Universidade Federal Fluminense, Av. Litor\^anea s/n, Gragoat\'a, Niter\'oi, R.J., 24210-340,                     Brazil 
                \and 
                R. Donangelo \at
                Instituto de F\'{\i}sica, Facultad de Ingenier\'{\i}a, C.C. 30, 11000 Montevideo, Uruguay
                \and
                W. Mittig \at
                National Superconducting Cyclotron Laboratory, Michigan State University, East Lansing, MI, USA}

\date{Received: date / Accepted: date}

\maketitle

\begin{abstract}

The scattering of identical nuclei at low energies exhibits conspicuous Mott oscillations which can be used to investigate the presence of 
components in the predominantly Coulomb interaction arising from several physical effects. It is found that at a certain critical value of the 
Sommerfeld parameter the Mott oscillations disappear and the cross section becomes quite flat. We call this effect Transverse 
Isotropy (TI). The critical value of the Sommerfeld parameter at which TI sets in is found to be $\eta_{c} = \sqrt{3s +2}$, 
where $s$ is the spin of the nuclei participating in the scattering.  No TI is found in the Mott scattering of identical Fermionic nuclei. The 
critical center of mass energy corresponding to $\eta_c$ is found to be $E_c$ = 0.40 MeV for $\alpha + \alpha$ (s = 0) , 1.2 MeV for 
$^{6}$Li + $^{6}$LI (s = 1) and 7.1 MeV for $^{10}$B + $^{10}$B (s = 3). We further found that the inclusion of the nuclear interaction induces a significant modification in the TI.  
We suggest measurements at these sub-barrier energies for the purpose of extracting useful  information about the nuclear interaction between light heavy ions. We also suggest extending the study of the TI to the scattering of identical atomic ions.

\end{abstract}

\section{Introduction}
\label{intro}

It is quite well known that the scattering amplitude describing the scattering of particles with a long-range interaction, such as the Coulomb one, composed of  a slowly varying amplitude and a possibly  rapidly varying phase. This results in a quantal cross section which is devoid of oscillations and which exactly coincides with the corresponding classical cross section. The reason for this is understood semi-classically, in the sense that in the evaluation of the quantal amplitude using the stationary phase method, one finds only one dominant contribution \cite{CH13}. The absence of oscillations in the scattering of non-identical particles with long range interaction, is changed when dealing with identical particles as the scattering amplitude in this case has to be symmetrized (bosons) or anti-symmetrized (fermions). In these cases one encounters rather strong oscillations in the cross section arising from the interference of the amplitude evaluated at an angle $\theta$ and the one at $\pi - \theta$. When the interaction is dominated by the Coulomb one, the resulting, Mott, cross section is oscillatory and can be used to assess the importance of other weaker interactions which may be present. Further it has been suggested \cite{FMG77, PiG81} that heavy-ion systems, such as $^{12}$C on $^{12}$C and $^{16}$O 
on $^{16}$O at sub-barrier energies, may exhibit deviation from a pure Mott scattering of identical bosons, owing to the underlying 
Fermi nature of the constituent nucleons. These authors, invoke the idea of parastatistics to quantify their suggestion. Experiments 
at Yale of $^{12}$C on $^{12}$C and $^{16}$O on $^{16}$O seem to exhibit such behavior \cite{Bro77}. More recently, other type 
of parastatistics based on the notion of ``quons", was advanced \cite{GrH99}. In this work, it is suggested that the constituent 
particles, the nucleons, are generalized Fermions, called quons with a label $q$. For $q = -1$ one has a Fermion, and $q= 1$ a 
Boson. The nuclei themselves acquire a $q$ value of $q_{\rm nucleus} = q_{\rm nucleon}^{A}$. With this prescription one may 
ormulate the Mott scattering in terms of the parastatistics parameter $q$, and this results in a similar cross section as that of 
\cite{FMG77, PiG81}.\\

Further, the quest for information about the short range nuclear interaction from elastic scattering data has been going on for a long time. 
This is even more challenging in the case of elastic scattering of heavy ions, where the long range Coulomb interaction is very 
important, especially at low energies. This fact prompted researchers to measure the cross section at higher energies, where the 
Coulomb effects are concentrated in the very small angular region around $\theta = 0$. Useful information were obtained about the 
nuclear interaction at these higher energies, especially in systems where nuclear rainbow dominates \cite{BrS97, HuM84,KOB07}. 
One may still wonder if low energy scattering could be used to obtain such information. In fact, it has been shown that several useful 
nuclear information can be obtained when the energy is below the Coulomb barrier and the cross section is predominantly 
Coulomb~\cite{LTB82, RKL82,HFF84}. Any deviation from the Coulomb interaction,  even if very small, may lead to 
measurable change in the characteristics of the Mott oscillations in the scattering of identical nuclei.  This fact lead, among other 
things, to test the existence of color Van der Walls force in the Mott scattering of  $^{208}$Pb + $^{208}$Pb~\cite{HLP90, VML93}. 
In this paper we propose to study a special feature of the Mott scattering to obtain information about the short range nuclear interaction, 
which would, in principle, complement  the information obtained at high energies. This special feature is the apparent disappearance 
of the oscillations at a certain critical value of the Sommerfeld parameter. Preliminary work on this has been done in \cite{CDH01} where
this feature of the cross section was called ``Transverse Isotropy" (TI). 
A recent experiment~\cite{ASR06} on $\alpha + \alpha$ Mott scattering seem to show this TI. In \cite{CHM14} further work on the TI was performed and compared to the data. It was 
demonstrated that the TI is quite sensitive to the presence of the nuclear interaction, making it an 
attractive venue to look for the latter. In this paper we give a reasonably detailed account of our work on the Transverse Isotropy \cite{CDH01, CHM14}.\\

The paper is organized as follows. In Section 2 we discuss the salient features of the scattering of identical Bosons and Fermions. In Section 3 we derive and discuss the Mott scattering cross section for any spin, and carefully analyze the interference effects. We also derive the condition for the occurrence of TI. In section 4 we include the nuclear potential in the predominantly Coulomb scattering problem and show how the short range interaction results in important qualitative change in the behavior of angular distribution. Whereas in the our Mott case, there is only one energy at which TI occurs, in the presence of the nuclear interaction we find at least two such energies. We confront our theoretical findings with the available data for the system $\alpha + \alpha$ (s = 0), and suggest measurements of the angular distribution for the higher spin systems, $^6$Li + $^6$LI (s = 1), and $^{10}$B + $^{10}$B (s = 3). In section 5 we present a brief discussion of Mott scattering and the TI in the case of the scattering identical atomic ions. Finally, in section 6 we present concluding remarks.

\section{Identical Particle Scattering}

In the scattering of identical particles with spin, s, the scattering amplitude must be properly symmetrized or anti-symmetrized in the
full spin and spatial spaces, depending on whether the particles are Bosons (integer s) or Fermions ( semi-integer s). Considering the
unpolarized spin case, the cross section must then be calculated as the average over spin orientations in the initial state and a sum over
total spin ($S = s + s$) projections, $m_S$, in the final state, 
\begin{equation}
\sigma(\theta) = \frac{1}{(2s+1)(2s +1)} \sum_{m_{\scriptscriptstyle S} = -2s}^{2s}\big|F_{m_S}(\theta)\big|^2,
\end{equation}
where $F_{m_S}(\theta)$ could be proportional to either  $f^{(+)}(\theta) = f(\theta) + f(\pi - \theta)$ or $f^{(-)}(\theta) = f(\theta) - f(\pi -\theta)$, depending on whether $m_S$ corresponds to even, or odd spin
wave function. For Bosons, even spin states are attached to even angular amplitude, $f^{(+)}(\theta)$, and odd spin states are attached to $f^{(-)}(\theta)$. The situation is reversed in the case of Fermions. It is now a simple matter to obtain the  following general expression,
\begin{equation}
\sigma(\theta) = a\ \big|f^{(\pm)}(\theta)\big|^2 + b\ \big|f^{(\mp)}(\theta)\big|^2,
\end{equation}
where $a$ and $b$ are spin-dependent factors given by,
\begin{equation}
a = \frac{s + 1}{2s + 1}
\end{equation}
and
\begin{equation}
b = \frac{s}{2s + 1} .
\end{equation}

The above expression for $\sigma(\theta)$ can be simplified to give,
\begin{equation}
\sigma(\theta) = \sigma_{inc}(\theta) \pm \frac{ \sigma_{int}(\theta)}{2s + 1} .
\end{equation}
The + and - signs stand for Bosons (integer s) and Fermions (semi-integer s), respectively, and  $\sigma_{inc}(\theta)$ is the incoherent cross section,
\begin{equation}
\sigma_{\rm inc}(\theta) = \big| f(\theta)\big|^2+\big|f\left(\pi-\theta \right)  \big|^2,
\end{equation}
and $\sigma_{int}(\theta)$ is the interference term,
\begin{equation}
\sigma_{\rm int}(\theta) = 2\,{\rm Re}\Big\{ f^\ast (\theta)\  f \left(\pi-\theta \right) \Big\}.
\end{equation}

In the case of scattering of identical particles with polarized spins of both target and projectile the cross section is simply given by,

\begin{equation}
\sigma(\theta) = \big|f(\theta) \pm f\left(\pi-\theta \right) \big|^2,
\end{equation}
giving,
\begin{equation}
\sigma(\theta) = \sigma_{inc}(\theta) \pm \sigma_{int}(\theta).
\end{equation}

Experimentally, it is quite difficult to perform measurements involving fully spin polarized target and projectile. Accordingly we opt to analyze here the less ambitious spin unpolarized 
case exemplified by the cross section given in Eq. (2).

\medskip

Note that the incoherent part of the cross section is positive-definite, whereas the interference term
may assume positive or negative values. Further,  since the scattering amplitude must have a continuous derivative with respect to
$\theta$ and the cross section is symmetric with respect to $\theta = 90^{\rm o}$,
$\sigma(\theta)$ must have a vanishing slope at this angle. It is instructive to calculate the cross section at $\theta = 90^{\rm o}$. From the general expression we find for Bosons,

\begin{equation}
\sigma(90^{\rm o}) = 4\ \frac{s +1}{2s + 1} \ \big|f(90^{\rm o})\big|^2 
\end{equation}
and for Fermions,
\begin{equation}
\sigma(90^{\rm o}) = 4\  \frac{s }{2s + 1}\  \big|f(90^{\rm o})\big|^2 
\end{equation}
These values correspond to extrema in the cross section, as the first derivative of the cross section at $\theta = 90^{\rm o}$ is zero.
To find wether the cross section has a maximum or a minimum at $\theta = 90^{\rm o}$, we should look at
the sign of the second derivative of the angular distribution at this angle. We find, for Bosons,
\begin{equation}
\sigma^{\prime\prime}(90^{\rm o}) =  8a\ {\rm Re} \Big\{ f^{\star}(90^{\rm o}) f^{\prime\prime}(90^{\rm o})\Big\} + 8b\ \big|f^{\prime}(90^{\rm o})\big|^2 \label{sdB}
\end{equation}
and for Fermions,
\begin{equation}
\sigma^{\prime\prime}(90^{\rm o}) = 8a\ \big|f^{\prime}(90^{\rm o})\big|^2 + 8b\ {\rm Re}\big\{ f^{\star}(90^{\rm o}) f^{\prime\prime}(90^{\rm o}) \big\} 
\label{sdF}
\end{equation}

\medskip

Clearly the determining factor in the sign of $\sigma^{\prime\prime}(90^{\rm o})$ is ${\rm Re}\left\{f^{\star}(90^{\rm o}) f^{\prime\prime}(90^{\rm o})\right\} $, which can be negative. If so, and remembering that $a > b$, we may claim, that for Bosons, $\sigma^{\prime\prime}(90^{\rm o}) < 0$ and thus $\sigma(90^{\rm o})$ is a maximum, while for Fermions, $\sigma^{\prime\prime}(90^{\rm o}) > 0$, and $\sigma(90^{\rm o})$ is a minimum. In the case of fully polarized spin of both projectile and target, $a = 1$ and $b = 0$, and $\sigma^{\prime\prime}(90^{\rm o})$, acquires the simple form,
\begin{equation} 
\sigma^{\prime\prime}(90^{\rm o}) = 8\ {\rm Re} \big\{f^{\star}(90^{\rm o}) f^{\prime\prime}(90^{\rm o}) \big\}
\end{equation}
and 
\begin{equation}
\sigma^{\prime\prime}(90^{\rm o}) = 8\ \big|f^{\prime}(90^{\rm o})\big|^2.
\end{equation}
More specific information about the unpolarized spin case considered above in Eqs. (21) and (22) requires detailed knowledge about the basic scattering amplitude, $f(\theta)$. A particularly interesting analytical case is that of a pure Coulomb scattering which results in a symmetry-based Mott cross section. We turn to this in the following.

\section{Mott scattering}

We begin by considering a simple scattering problem: the collision of structureless
particles, interacting only through point-charge Coulomb forces. In this case, the
problem has an analytical solution (see, e.g. Ref.~\cite{CH13}),
\begin{equation}
f_{\rm \scriptscriptstyle{C}}(\theta)=-\frac{a}{2}\  e^{2i\sigma_{0}}\ \
\frac{e^{-i\eta\ln(\sin^{2}\theta/2)}}{\sin^{2}(\theta/2)} ,
\label{fc}
\end{equation}
where $\eta$ and $a$ are respectively the Sommerfeld parameter and half the distance
of closest approach in a head-on collision, given by
\begin{equation}
\eta = \frac{q^2}{\hbar v},\qquad a=\frac{q^2}{2E}.
\end{equation}
Above, $q$ is the charge of the identical particles (the projectile and the target),
$v$ is the relative velocity, and $\sigma_{0}$ is the s-wave Coulomb phase shift
\begin{equation}
\sigma_{0}=\arg \big\{ \Gamma(1+i\eta) \big\},
\label{sig0}
\end{equation}
\smallskip

\noindent with $\Gamma$ standing for the usual Gamma-function.

\medskip

The Coulomb amplitude, Eq.~(\ref{fc}), contains a single contribution, and accordingly the cross section associated with it, the Rutherford cross section, $\sigma_{R}(\theta) = (a/4\sin^4{\theta/2})$, is structureless. Further, it coincides exactly with the classical cross section. As mentioned in the introduction, this fact is a direct consequence of the fact that the phase of the corresponding partial wave $S$-matrix \cite{CH13}, $S_l \approx \exp{2i[\sigma_0 + (l + 1/2)\tan^{-1}{[\eta/(l + 1/2)]} + 1/2\eta \ln{[(l + 1/2)^2 + \eta^2]}}$, is a slowly varying function of $l+1/2$, whose derivative with respect to the angular momentum, namely the deflection function, is a monotonic function of $l+1/2$, $\Theta(l) = 2 d\sigma_l/d(l+1/2) = \eta/(l + 1/2)$. The amplitude is given by an infinite, divergent, $l$ sum. However, when use is made of the stationary phase approximation, then the $l$-sum, which is conveniently replaced by an integral over the continuous variable $l$+ 1/2, can be evaluated and the result is just Eq.~(\ref{fc})!. In general, the deflection function is not a monotonic function of $l$, and the resulting amplitude will be a sum of a few complex contributions. The resulting cross section  would then exhibit an interference pattern. In summary, the Coulomb amplitude  describing the scattering of a structures charge, is just the product $f_{\rm \scriptscriptstyle{C}}(\theta) = A(\theta) e^{i\phi (\theta)}$, with the amplitude $A$ and phase $\phi$ being real functions of $\theta$, given in Eq.~(\ref{fc}). \\

We can write for the Mott cross section,
\begin{equation}
 \sigma_{\scr M}(\theta) =  \sigma_{\rm inc}(\theta) \pm \frac{1}{2s + 1}\sigma_{\rm int}(\theta),
\label{inc-int_bar}
\end{equation}
with
\begin{equation}
\sigma_{\rm inc}(\theta )=\frac{a^2}{4}\,\left[ \frac{1}{\sin ^{4}\left( \theta /2\right) }+\frac{1}{\cos ^{4}\left( \theta /2\right) }\right]   \label{siginc}
\end{equation}
and
\begin{equation}
\sigma_{\rm int}(\theta ) =\frac{a^2}{4}\,\left[ 2\frac{\cos \left[ 2\eta \ln \left( \tan (\theta/2)\right) \right] }{\sin ^{2}(\theta /2)\,\cos ^{2}(\theta /2)}\right]  \label{sigint}.
\end{equation}

\medskip

It is interesting to exhibit the full $\eta$ dependence of the Mott cross section. For this purpose we write $a^2 /4 = \left[\hbar^2 /(q^2 A\, m_0)\right]^2 \eta^4$, where $m_0$ is the average nucleon mass. The Mott cross section at $\theta = 90^{\rm o}$ is then $\sigma_{\scr M}(\theta = 90^{\rm o}) = 16(s + \delta_{stat})/(2s + 1) \left[\hbar^2 /(q^2 A\, m_0)\right]^2 \eta^4$. Here $\delta_{stat}$ = 1 or 0, depending whether the system is Bosonic or Fermionic, respectively. Dividing by the $\eta$- and $s$-independent factor of $\sigma_{\scr M}(\theta = 90^{\rm o})$,  Eqs.~(\ref{siginc}) and (\ref{sigint}) then become,
\begin{equation}
\overline {\sigma}_{\rm inc}(\theta) = \frac{1}{16\left[\hbar^2 /(q^2 A\, m_0)\right]^2}\sigma_{\rm inc}(\theta) = \frac{1}{16}\eta^4 \left[ \frac{1}{\sin ^{4}\left( \theta /2\right) }+\frac{1}{\cos ^{4}\left(\theta /2\right)}\right]  ,
\label{scsiginc}
\end{equation}
\begin{equation}
\overline {\sigma}_{\rm int}(\theta) = \frac{1}{16\left[\hbar^2 /(q^2 A\, m_0)\right]^2}\sigma_{\rm inc}(\theta) = \frac{1}{16}\eta^4 \left[ 2\frac{\cos \left[ 2\eta \ln \left( \tan(\theta/2)\right) \right] }{\sin ^{2}(\theta /2)\,\cos ^{2}(\theta /2)}\right] 
\label{scsigintB}
\end{equation}

\medskip

The condition for a maximum or a minimum at $\theta = 90^{\rm o}$ in the Mott cross section above can be easily found by calculating the second derivative of the scaled cross sections of Eqs.~(\ref{scsiginc}), and ({\ref{scsigintB}), which, for Bosons, come out to be
\begin{equation}
\overline \sigma^{\prime\prime}_{\scr M}(\theta)\, \big|_{\theta = 90^{\rm o}}  = \left[ \frac{s + 1}{2s + 1} \right]\ \eta^{4}\big[-\eta^2 + 3s +2\big] 
\label{scsdsigB}
\end{equation}
and for Fermions,
\begin{equation}
\overline \sigma^{\prime\prime}_{\scr M}(90^{\rm o})\, \big|_{\theta = 90^{\rm o}}  = \left[ \frac{s}{2s + 1} \right] \ \eta^{4}\big[\eta^2 + 3s +1]\big] .
\label{scsdsigF}
\end{equation}

\medskip

The above scaled second derivative of the Mott cross section for Bosons, rises from zero value at $\eta = 0$, attaining a maximum value of  $(4/3)^3\sqrt{(3s+2)^3/(2s+1)}$ at $\eta_m = \sqrt{2(3s+2)/3}$, and then it dives to zero at $\eta_c = \sqrt{3s + 2}$, followed by a steep delve into negative values. On the other hand, for Fermions with $s$ = 0, the Mott cross section is zero at $\theta = 90^{\rm o}$, and so is its, second derivative. On the other hand, for $s >$1 the Fermion Mott cross section will have a non-zero value, $\sigma(90^{\rm o}) = 16[s/(2s +1)]\left[\hbar^2 /(q^2 A\, m_0)\right]^2 \eta^4$,  and its second derivative at $\theta = 90^{\rm o}$, given by Eq. (\ref{scsdsigF}) multiplied by the factor $16\left[\hbar^2 /(q^2 A\, m_0)\right]^2$, is still positive indicating a minimum.\\

From the above analysis, it is clear that for Bosons, the second derivative is negative for $\eta > \sqrt{3s + 2}$, and positive for $\eta < \sqrt{3s + 2}$, while for Fermions, it is always positive. For large values of $s$, the second derivative is always positive as the cross section in this case is purely classical. It is interesting to consider the condition that the cross section has an inflection point at $\theta = 90^{\rm o}$. This implies a zero second derivative, which results in a critical value of $\eta$ for Bosons, $\eta_{c} = \sqrt{3s + 2}$, while no such inflection point exists for Fermions. The Bosonic system at an energy corresponding to $\eta_{c}$ exhibits a flat angular distribution as if the quantum interference is washed out by the the interaction. We see the natural separation into regions of minima and maxima. We have examined this phenomenon in a previous papers \cite { CDH01}, and have called the effect \textit{Transverse Isotropy} (TI). We illustrate this behavior showing in Fig.~\ref{LiLi} for s = 1, the normalized Mott cross sections  for a value
of the Sommerfeld parameter below $\eta_0$,  ($\eta = 0.2$) and for one value above ($\eta = 4.0$). As 
expected the former has a minimum at $\theta = 90^{\rm o}$, whereas the latter has a maximum. However, the 
most interesting feature of this figure is the TI which removes the interference effects in the cross section at the critical value of the Sommerfeld parameter,
$\eta_{c} =\sqrt{5}$. In this case, the cross section is remarkably flat around $90^{\rm o}$. 
\medskip

\begin{figure}
\centering
\includegraphics[width=8 cm]{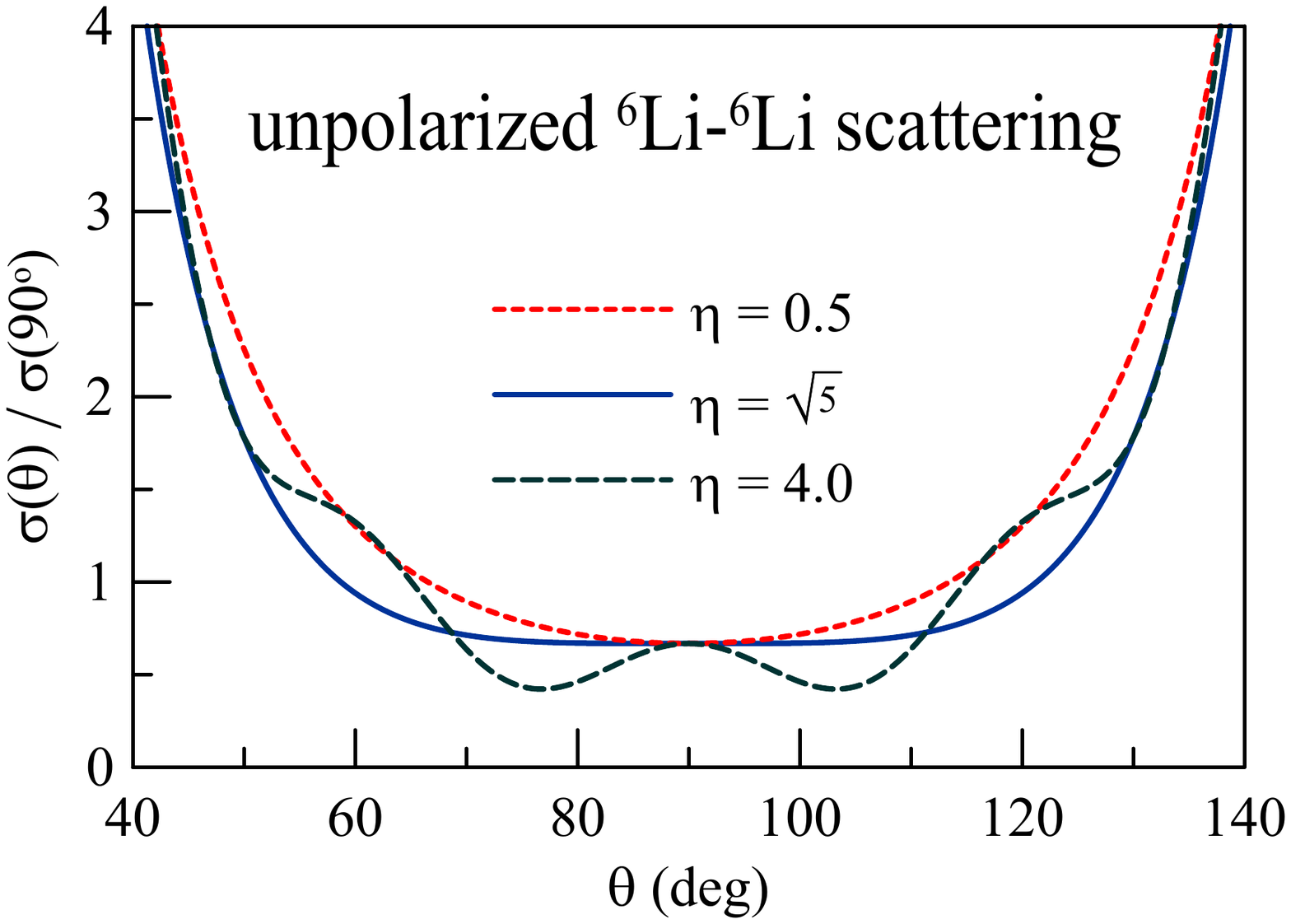}
\caption{(Color online) Mott cross sections for the system $^6$Li + $^6$Li  (s = 1)at three values of the Sommerfeld parameter.
The Mott cross sections are normalized with respect to their values at $\theta = 90^{\rm o}$. The normalized cross sections are multiplied by the spin factor $(s + 1)/(2s + 1)$ }
\label{LiLi}
\end{figure}

\medskip

\section{Transverse isotropy in Nuclear Physics}

In principle, nuclear systems can be good candidates to exhibit the flat cross sections discussed in the previous
section. However, nuclei interact both through Coulomb and nuclear forces. Thus, the prediction of a flat cross 
section at the critical value of the Sommerfeld parameter will only be valid if the corresponding collision energy
is below the height of the Coulomb barrier. We should then check if there are nuclear systems satisfying this
condition. As qualitative approach to the problem, we set $Z_{\scr P}=Z_{\scr T}=Z$ and
$A_{\scr P}=A_{\scr T}=2Z$, evaluate the collision energy corresponding to $\eta_0$, 
\[
E_0 =\frac{Z(Z^2 e^2)^{2}\,m_0}{2\hbar^{2} \eta_{0}^2} = \frac{0.025\ Z^5}{3s + 2}\ [{\rm MeV}]
\]
where $m_0$ is the unit mass of the nucleon ($m_{0} c^2 \simeq$ 931 MeV)
and estimate the barrier height by the approximate expression,
\[
V_{\scr B} \simeq \frac{Z_{\scr P}  Z_{\scr T} e^2}{r_0\left( A_{\scr P}^{1/3}+A_{\scr T}^{1/3} \right)}
=\frac{e^2}{r_0}\ Z^{5/3}\,2^{-4/3} = 0.48\, Z^{5/3}\ [{\rm MeV}]
\]
We now construct the ratio $E_0/V_{\scr B}$, and use for $\eta_{0}^2$ = 3s + 2, to get

\begin{equation}
\frac{E_0}{V_{\scr B}} = \frac{2^{1/3}\,Z^{10/3}e^{2}\,m_{0}\,r_0}{\hbar^{2}(3s + 2)}= \frac{0.052\ Z^{10/3}}{3s + 2},
\end{equation}
where we have used $r_0$ = 1.2 fm, Thus for $^{2}$H (s = 1) we get 0.01, for $^4$He (s = 0), 0.26, for $^6$Li (s = 1), 0.405, and for $^{10}$B (s = 3), 1.01.
The nucleus-nucleus interaction at low energies can be generally represented by the following
\begin{equation}
V(r) = \frac{q^2}{r} + V_{\scr N}(r) + \sum_{i} \Delta V_i,
\end{equation}
where $V_{\scr N}(r)$ is the short range nuclear interaction to be discussed in the following, and the sum runs over small contributions arising from several physical effects, such as vacuum polarization, electron screening, nuclear polarizations, etc. These small effects were studied in \cite{RKL82,HLP90,VML93}. In the following we restrict ourselves to the study of the influence of $V_{\scr N}(r)$ on the TI.\\

To investigate the effect of the nuclear interaction on the TI, we use the the modified amplitude obtained using the stationary phase method,
\begin{equation}
f(\theta) = - \frac{a}{2}\ e^{2i\sigma_0}\ \frac{e^{-i\eta\ln{\sin^2{\theta/2}} + i\Delta_{\scr N}}}{\sin^2{\theta/2}}
\end{equation}
where the nuclear phase $\Delta_{\scr N}(\theta)$ given to first order in the nuclear interaction, $V_{\scr N}(r)$, is
 \begin{equation}
 \Delta_{\scr N} = -\, \sqrt{\frac{2\mu}{\hbar^2}}\ \int_{r_0}^{\infty}\frac{V_{\scr N}(r)}{\sqrt{E - V_{\scr C}(r) - \frac{\hbar^2 l(l + 1)}{2\mu r^2}}}\ dr = -\frac{2}{\hbar}\int_{0}^{\infty}V_{\scr N}(r(t))\ dt. \label{deltaN}
\end{equation} 
\noindent
Above,  $r_0$  is the distance of closest approach, and we have changed variable from $r$ to $t$ through the classical relation $dr = v\, dt = \sqrt{2/ \mu}\sqrt{E - V_{\scr C}(r) - \hbar^2l(l + 1)/(2\mu r^2)} dt$. We take for the nuclear interaction a conventional Saxon-Woods form,
\begin{equation}
V_{\rm N}(r)= \frac{V_0}{1+\exp\left[ \left(r-R_0\right)/a_0  \right]}.
\end{equation}
At the low energies being considered here, the above potential is felt at large distances, where it behaves as an exponential, $V_{\scr N} \approx V_0 \exp{[(R - r)/a_0]}$. 

\smallskip

Expanding $r(t)$ around $t = 0$, we get 
\[
r (t) = r_0 - \frac{1}{2}\ r^{\prime\prime}(0)\ t^2,
\]
where $r^{\prime\prime}(0)$ is the radial acceleration at closest approach, and  the integral in the nuclear phase can be evaluated in closed from. We get
\begin{equation}
\Delta_{\scr N} = -\frac{2}{\hbar} \ \sqrt{\frac{\pi a}{r^{\prime\prime}(0)}}\ V_{0}\ e^{(R - r_0)/a_0} .
\end{equation}
The angle-dependence in $\Delta_{\scr N}$ resides primarily in the exponential $\exp\left(-r_0/a\right)$, with $r_0 = a + \sqrt{a^2 + b^2}$, where $b$ is the impact parameter, given by the Coulomb relation, $b(\theta) = a \cot{\theta/2}$. The second derivative of the cross section can be evaluated easily, using the general formula of Eq.~(\ref{sdB}),
\begin{equation}
\sigma^{\prime\prime}(90^{\rm o}) =  8\ \frac{s +1}{2s +1}\  {\rm Re}\left\{f^{\star}(90^{\rm o}) f^{\prime\prime}(90^{\rm o})\right\} + 
8\ \frac{s}{2s +1}\ \big|f^{\prime}(90^{\rm o})\big|^2 .
\label{sdB1}
\end{equation}
giving,
\begin{equation}
\overline{\sigma}^{\prime\prime}(90^{\rm o}) \approx \frac{s + 1}{2s + 1}\ \eta^4 \ \Big[-\big(\eta - \Delta^{\prime}_{\scr N}(90^{\rm o})\big)^2 + (3s + 2)\Big].\label{sdn}
\end{equation}

It is clear that the inclusion of the nuclear interaction affects the value of $\eta$ at which the TI sets in. This includes a simple shift of value, as well as a possible two or more solutions for the critical $\eta_c$. Numerical calculation of this effect using the exact solution of the Schr\"dingier equation with the Aky\"uz-Winther nuclear interaction \cite {AkW81} performed in \cite{CHM14} for the system $\alpha + \alpha$ has shown such a behavior, with the lowest critical energy at $E_0 = 0.4$ MeV, and the highest one at center of mass energy $E_0 = 1.10$ MeV.  This seems to be in qualitative agreement with the data of \cite{ASR06}. In fact the latter data exhibits TI at three critical energies. The lowest critical energy shifts slightly, taking the value $E_0=0.47$ MeV (as compared to the theoretical value 0.4 MeV) and the second one moves to $E_0^\prime = 2.41$ MeV. Between these two 
energies the second derivative remains very small, getting very close to zero around $E_{c.m.}$  = 2.2 MeV. 
This behavior is potentially useful for obtaining information about the nuclear interaction at sub-barrier energies. The behavior of the loosely bound system $^6$Li + $^6$Li  at an energy in the vicinity of the critical energy $E_0$ = 1.2 MeV, should be similar. We encourage experimentalists to investigate the TI in this system and possibly the $^{10}$B + $^{10}$B (s = 3) system, with a critical energy, $E_0 = 7.1$ MeV,  in order to obtain invaluable information about the short- ranged nuclear interaction at the long distances reached at these low energies.

\section{Transverse Isotropy in atomic  ion-ion systems}
In the scattering of identical atomic ions, the dominant part of the interaction is still the point Coulomb $1/r$ potential. Thus the cross section should be the Mott one
modified by the spin of the ions as discussed above. There are other important components in the ion-ion interaction which have to be considered, such as the attractive monopole-dipole $1/r^4$ polarization contribution and the repulsive Casimir-Polder, $1/r^5$ contribution \cite{bab10,FeS79,CaP48},
\begin{equation}
V_{\scr A}(r) = \frac{q^2}{r} - 2\ \frac{1}{2}\ \frac{q^2 \alpha(o)}{r^4} + 2\ \frac{11}{4\pi}\ \lambda_{\scr C}\frac{q^{2}\alpha(o)}{r^5} + \cdots
\end{equation}
where $\alpha(0)$ is the static dipole polarizability of the ion and $\lambda_{\scr C} = \hbar/m_{e}c$ is the Compton wave length. 

\smallskip

The $r^{-4}$ and $r^{-5}$ components will certainly affect our findings about the Transverse Isotropy in the angular distribution. In fact the last equation of the last section, Eq.~(\ref{sdn}) is directly applicable to the current case of atomic ion-ion scattering,
\begin{equation}
\overline{\sigma}^{\prime\prime}(90^{\rm o}) \approx \frac{s + 1}{2s + 1}\ \eta^4 \ \Big[-\big(\eta - \Delta^{\prime}_{\scr A}(90^{\rm o})\big)^2 + (3s + 2)\Big].\label{sda}
\end{equation}

where the atomic  ion-ion phase, $\Delta^{\prime}_{\scr A}(90^{\rm o})$ is calculated as before, Eq.~(\ref{deltaN}),
\begin{equation}
  \Delta_{\scr A} = -\, \sqrt{\frac{2\mu}{\hbar^2}}\, \int_{r_0}^{\infty}\frac{V_{\scr A}(r) -\frac{q^2}{r}}{\sqrt{E - V_{\scr C}(r) - \frac{\hbar^2 l(l + 1)}{2\mu r^2}}}\ dr = -\frac{2}{\hbar}\int_{0}^{\infty}[V_{\scr A}(r(t)) -\frac{q^2}{r(t)}]\ dt. \label{deltaA}
\end{equation}

Thus,

 \begin{equation}
  \Delta_{\scr A} = -\, \frac{2}{\hbar}\int_{0}^{\infty} \big[- \frac{q^2 \alpha(o)}{r(t)^4} + \, \frac{11}{2\pi}\ \lambda_{\scr C}\frac{q^{2}\alpha(o)}{r(t)^5}\big]\ dt. \label{sdA}
\end{equation}

This last equation can be evaluated in a straightforward fashion, and its second derivative with respect to $\theta$ at $90^{\rm o}$, Eq.~(\ref{sda}), can be also easily computed. The condition for the occurrence of TI is then obtained as the solution of the equation, $\eta^4\big[-\big(\eta -  \Delta^{\prime}_{\scr A}(90^{\rm o})\big)^2 + (3s + 2)\big] = 0$. The full details of the TI in the identical  atomic ion-ion scattering case, will be discussed elsewhere. 

\section{Conclusions}
In this contribution we have summarized and extended the development of the phenomenon of Transverse Isotropy predicted to exist in the Mott scattering of identical charged Bosonic particles. We have derived an approximate formula which exhibits the influence of the nuclear interaction on the  critical value of critical value of the Sommerfeld parameter, $\eta_c$, at which the TI sets in.  Exact numerical evaluation of the TI using the the Aky\"uz-Winther interaction are presented for the $\alpha + \alpha$ system and compared to the data of \cite{ASR06} as done in \cite{CHM14}. The system $^6$Li + $^6$Li is also briefly discussed. Finally application of the theory to the scattering of atomic ions is proposed.

\bigskip

\end{document}